\documentclass[
twocolumn,
twoside,
 amsmath,amssymb,
 aps,prd]{revtex4-2}

\usepackage{graphicx,dcolumn,bm,subfigure}

\begin{document}

\preprint{MUPB/Conference section: }

\title{Neutrino oscillations in gravitational fields and astrophysical applications}

\author{Maxim Dvornikov}
\affiliation{Pushkov Institute of Terrestrial Magnetism, Ionosphere and Radiowave Propagation (IZMIRAN), 108840 Moscow, Troitsk, Russia}
\email{maxdvo@izmiran.ru}

\date{\today}

\begin{abstract}
The neutrino propagation and oscillations in various gravitational fields are studied. First, we consider the neutrino scattering off a black hole accounting for the neutrino spin precession. Then, we study the evolution of flavor neutrinos in stochastic gravitational waves. The astrophysical applications of the obtained results are considered.
\end{abstract}
                     
\maketitle

Oscillations of neutrinos, confirmed experimentally, allow one to explore physics beyond the standard model. This phenomenon can take place only for nonzero masses and mixing between different neutrino generations. We also mention the possibility of transitions between active left handed and sterile right handed neutrinos. This process is called neutrino spin oscillations.

The interaction with external fields can modify the process of neutrino oscillations. Gravitational fields, in spite of their weakness, can also affect neutrino oscillations. We studied neutrino spin oscillations in background matter under the influence of an electromagnetic field in static curved spacetimes in Refs.~\cite{Dvo06,Dvo13}. Spin oscillations of neutrinos in their scattering off a black hole (BH) were analyzed in Refs.~\cite{Dvo19c,Dvo20a,Dvo21a}. This research was inspired by the recent observation of the event horizon silhouette of a supermassive BH (SMBH) in the center of M87. An accretion disk around BH can be a source of both photons and neutrinos. Neutrino spin oscillations in a gravitational wave (GW) were considered in Ref.~\cite{Dvo19a}. We examined neutrino flavor oscillations in stochastic GWs in Refs.~\cite{Dvo19b,Dvo20b,Dvo21b}. The study of the neutrino interaction with GWs was motivated by the direct detection of GWs in terrestrial experiments. In the present work, we summarize our results in Refs.~\cite{Dvo21a,Dvo21b}.

First, we study neutrino spin oscillations in the particle scattering off a rotating BH surrounded by a dense magnetized accretion disk. The invariant neutrino spin vector $\bm{\zeta}$, which is defined in the rest frame of the locally Minkowskian
frame, obeys the precession equation~\cite{Dvo13,Dvo21a}, $\dot{\bm{\zeta}} = 2 (\bm{\zeta}\times\bm{\Omega})$. The components of the vector $\bm{\Omega}$ incorporate the contributions of all the external fields. They are given in the explicit form in Ref.~\cite{Dvo21a}. Note that one has to account for both the neutrino spin precession and the evolution of the neutrino velocity to describe the neutrino polarization. We suppose that neutrinos move along the geodesic lines in their scattering off BH.

We use the Kerr metric for the description of the spacetime around a rotating SMBH with $M\sim10^{8}\,M_{\odot}$, which is surrounded by a relativistic accretion disk. The electron number density in the disk scales as $n_{e}(r)\propto n_{e}^{(0)}r^{-3/2}$, where $n_{e}^{(0)}\sim10^{18}\,\text{cm}^{-3}$ is the number density in the vicinity of SMBH. We suppose that the magnetic field in the disk is poloidal and its strength scales as $B(r) \propto B_{0}r^{-5/4}$. The magnetic field  in the vicinity of BH is taken to be $B_{0}=3.2\times10^{2}\,\text{G}$. Neutrinos are supposed to be Dirac particles having the anomalous magnetic moment $\mu = 10^{-14}\mu_\mathrm{B}$, where $\mu_\mathrm{B}$ is the Bohr magneton. The matter of the accretion disk is taken to rotate around SMBH on circular orbits, which are the realistic geodesics for plasma particles forming this disk. Note that an accretion disk can both corotate and counter-rotate BH. We suppose that neutrinos move in the equatorial plane only.

\begin{figure*}
  \centering
  \subfigure[]
  {\label{fig:spina}
  \includegraphics[scale=.3]{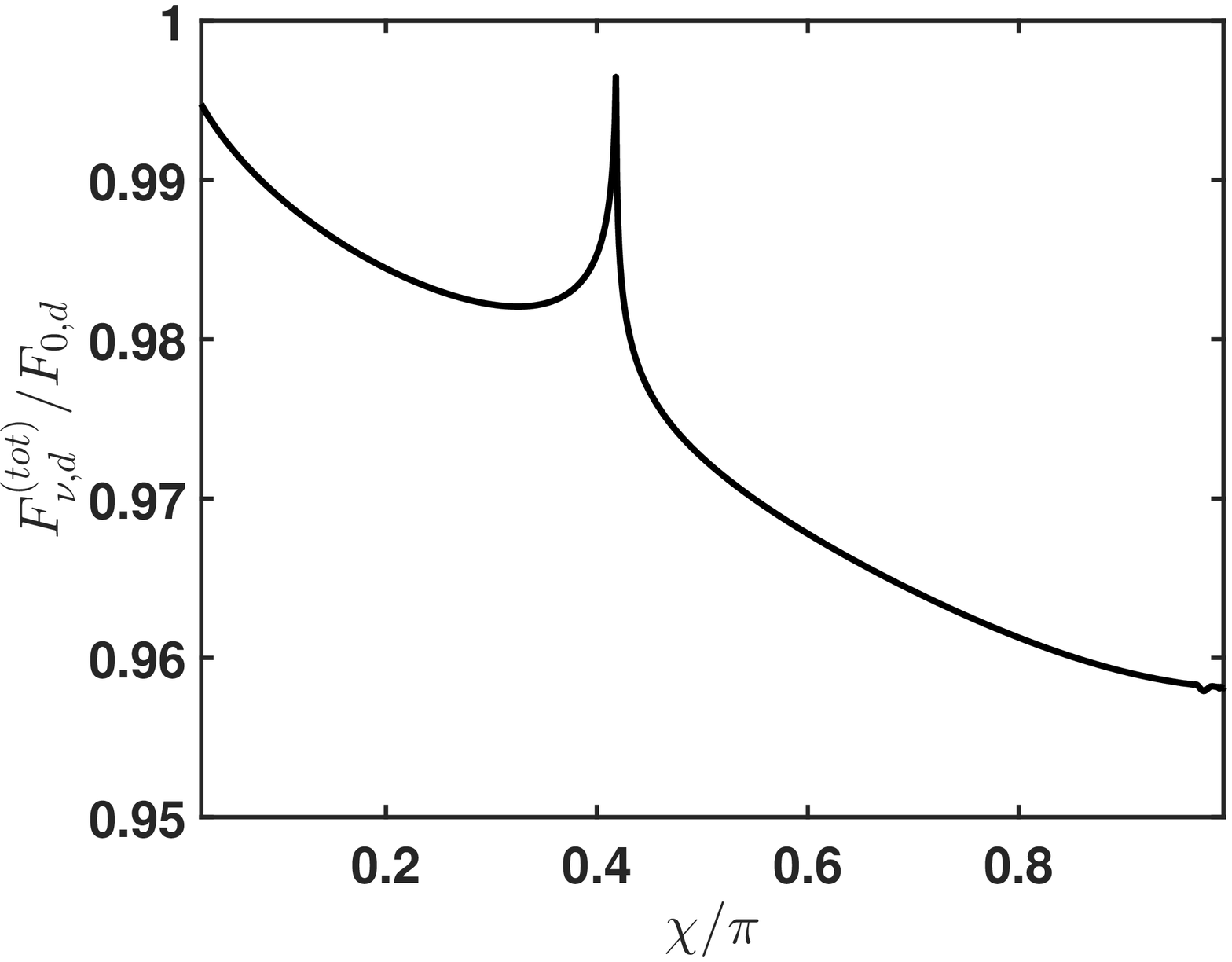}}
  \hskip-.6cm
  \subfigure[]
  {\label{fig:spinb}
  \includegraphics[scale=.3]{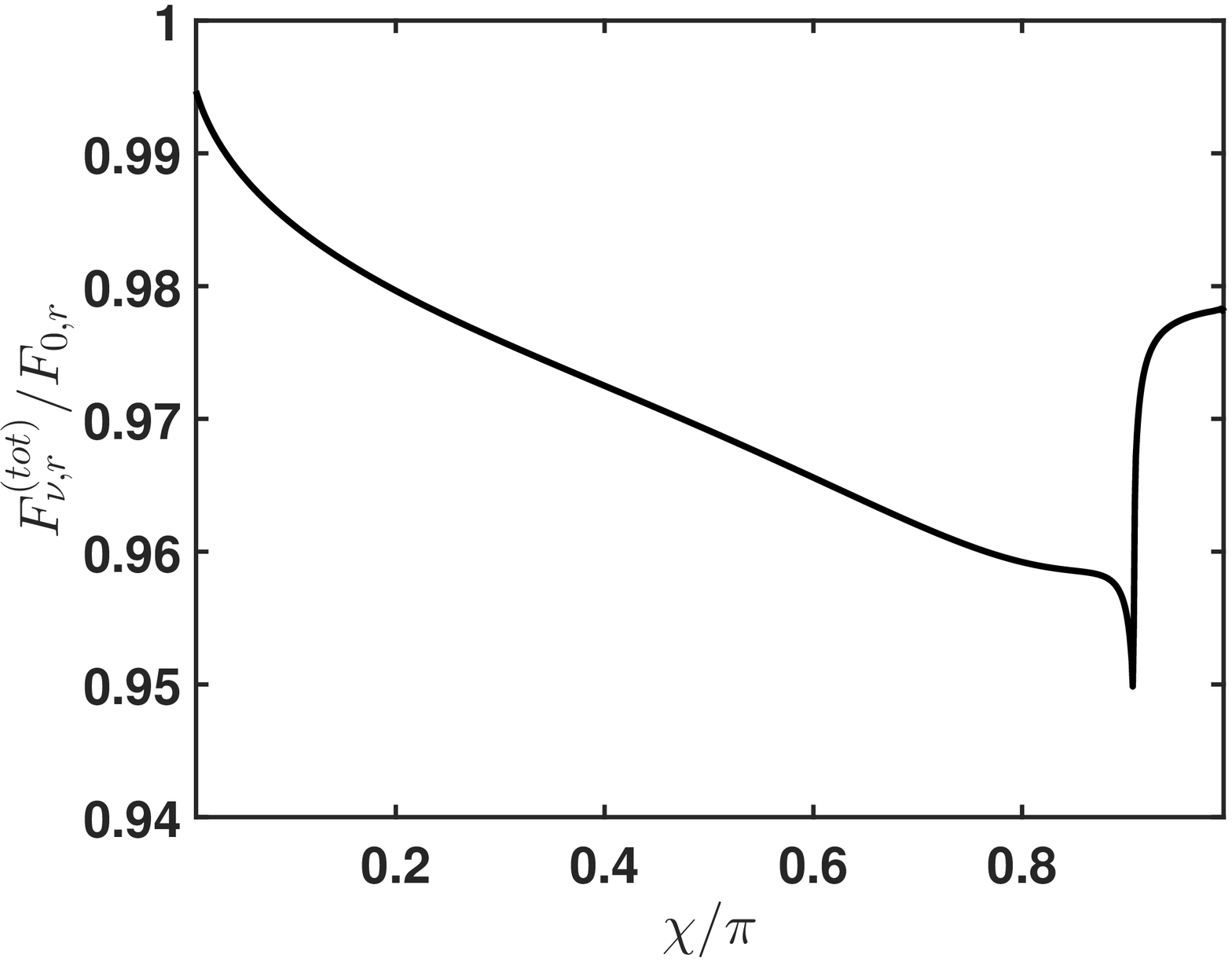}}
  \protect
  \caption{(a) The total flux $F_{\nu,d}^{(\text{tot})}$, normalized by the flux of scalar particles $F_{0,d}$,
  for the direct scattering (a detector is bent towards the BH rotation)
  for BH with the angular momentum $J = M^2$ versus the scattering angle $\chi$.
  (b) The total flux $F_{\nu,r}^{(\text{tot})}$ for the retrograde scattering (a detector is bent in the opposite direction with respect to the BH rotation)
  for BH with $J = 0.2 M^2$.  The accretion disk counter-rotates BH in both panels. These figures are taken from Ref.~\cite{Dvo21a}.\label{fig:spin}}
\end{figure*}

The total neutrino fluxes scattered off BH accounting for the spin evolution for different angular momenta of BH are shown in Fig.~\ref{fig:spin}. One can see in Fig.~\ref{fig:spin} that there are spikes in the fluxes at certain scattering angles. The appearance of these spikes can be accounted for qualitatively by the neutrino interaction with moving matter. In the neutrino gravitational scattering, there are situations when the neutrino velocity has a nonzero angle with respect to the plasma velocity. In this case, a neutrino spin-flip can happen.

One can see in Fig.~\ref{fig:spin} that the observed neutrino flux can be suppressed by several percent because of spin oscillations in the neutrino gravitational scattering. The major reduction of the flux happens for the backward neutrino scattering when $\chi = \pi$. We hope that the predicted effect can be observed by current or future neutrino telescopes.

Now, we turn to the problem of the propagation and flavor oscillations of supernova (SN) neutrinos under the influence of stochastic GWs. The flavor content of a neutrino beam can be described by the $3\times 3$ density matrix $\rho_f$ in the flavor basis. We can define the new density matrix $\rho' = \exp(\mathrm{i}H_m^{(\mathrm{vac})}t) U^\dagger \rho_f U \exp(-\mathrm{i}H_m^{(\mathrm{vac})}t)$, where $(H_m^{(\mathrm{vac})})_{ab} = \tfrac{m_a^2}{2E}\delta_{ab}$ is the vacuum oscillations Hamiltonian in the mass basis corresponding to the masses $m_a$, $E$ is the mean neutrino energy, and $U$ is the mixing matrix. In the presence of stochastic GWs, the averaged density matrix $\langle \rho' \rangle$ was shown in Ref.~\cite{Dvo21b} to obey the equation 
$
  \tfrac{\mathrm{d}}{\mathrm{d}t}
  \left\langle
    \rho'
  \right\rangle (t)=
  g(t)[H_m^{(\mathrm{vac})},[H_m^{(\mathrm{vac})},
  \left\langle
    \rho'
  \right\rangle (t)]]
$,
where the function $g(t)$ contains the correlators of the amplitudes $h_{+,\times}$ of the `plus' and `cross' polarizations of GW.

\begin{figure}
  \centering
  \includegraphics[scale=.36]{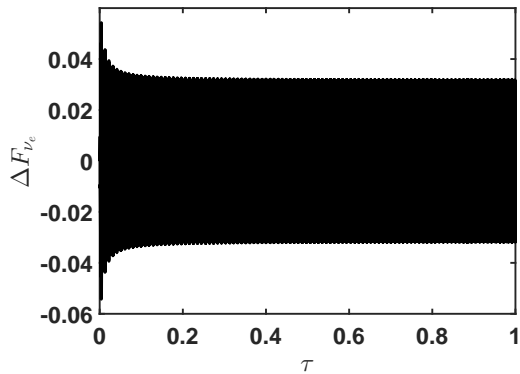}
  \protect
  \caption{(a) The correction the electron neutrino flux $\Delta F_{\nu_{e}}\propto\Delta P_{\nu_{e}}$,
  where $P_{\nu_{e}}$ is the probability to detect $\nu_{e}$,
  owing to the neutrino interaction with stochastic GWs. The parameters
  of the system are $\Delta m_{21}^{2}=7.5\times10^{-5}\,\text{eV}^{2}$,
  $E=10\,\text{MeV}$, $\theta_{12}=0.6$,
  $\theta_{23}=0.85$, $\theta_{13}=0.15$, $\delta_{\mathrm{CP}}=3.77$, $L=10\,\text{kpc}$ is
  the maximal propagation distance, $\Omega_{0}=10^{-9}$ is the maximal normalized spectral energy density of GWs,
  and $f_{\mathrm{min}}=10^{-10}\,\text{Hz}$ is the minimal frequency of the GW spectrum.
  The normal neutrino mass ordering is adopted. This figure is taken from Ref.~\cite{Dvo21b}.\label{fig:deltaF}}
\end{figure}

We take that a beam of neutrinos is emitted in a core-collapsing SN, which can be considered as almost a point-like source. In this case, only solar neutrinos oscillations channel has a nonvanishing contribution to the observed fluxes. We also suppose that these SN neutrinos interact with stochastic GWs emitted by merging SMBHs. The equation for $\langle \rho' \rangle$ can be solved analytically in this situation. The behavior of the GW contribution to the flux of electron neutrinos is shown in Fig.~\ref{fig:deltaF}. In Fig.~\ref{fig:deltaF}, we demonstrate the evolution of $\Delta F_{\nu_e} = F_{\nu_e} - F^{(\text{vac})}_{\nu_e}$ along the neutrino propagation distance $x = \tau L$. Here $F_{\nu_e}$ is the flux accounting for the GW contribution and $F^{(\text{vac})}_{\nu_e}$ is the flux corresponding to vacuum oscillations. The fluxes of $\nu_\mu$ and $\nu_\tau$ are considered in Ref.~\cite{Dvo21b}.

One can see in Fig.~\ref{fig:deltaF} that the interaction of SN neutrinos with stochastic GWs can modify the observed flux, e.g., of electron neutrinos by a few percent. It means that one expects the change of the
SN neutrinos fluxes by $\sim\pm350$ events, in case of the Super-Kamiokande,
and by $\sim\pm3750$ events, for the Hyper-Kamiokande if a SN explosion happens in our Galaxy.

\end{document}